\documentclass[aps,prl,showpacs,nofootinbib,twocolumn,superscriptaddress]{revtex4}

\usepackage[dvips]{graphicx}
\usepackage{amssymb,amsmath,amstext,amsthm,amsfonts}
\usepackage{slashed}

\begin{document}

\newcommand{\reffig}[1]{Fig.~\ref{#1}}
\newcommand{\refeq}[1]{Eq.~(\ref{#1})}
\newcommand{\refcite}[1]{Ref.~\cite{#1}}
\newcommand{\refscite}[1]{Refs.~\cite{#1}}
\newcommand{\refetal}[1]{\emph{et~al.}~\cite{#1}}
\newcommand{\dd}{\mathrm{d}}
\newcommand{\kz}{k^0}
\newcommand{\pz}{p^0}
\newcommand{\subtot}{\text{tot}}
\newcommand{\res}[3]{#1$_{#2}$(#3)}
\newcommand{\lep}{\ell}
\newcommand{\myvec}[1]{\boldsymbol{\mathrm{#1}}} 
\newcommand{\myunit}[1]{\mbox{$\,\text{#1}$}}
\newcommand{\GeV}{\myunit{GeV}}

\renewcommand{\theenumi}{\arabic{enumi}}
\renewcommand{\labelenumi}{(\theenumi)}

\newcommand{\plotscale}{.80}

\title{Neutrino induced pion production at MiniBooNE and K2K}

\author{T.~Leitner}
\email{tina.j.leitner@theo.physik.uni-giessen.de}
\affiliation{Institut f\"ur Theoretische Physik, Universit\"at Giessen, Germany}

\author{O.~Buss}
\affiliation{Institut f\"ur Theoretische Physik, Universit\"at Giessen, Germany}

\author{U.~Mosel}
\affiliation{Institut f\"ur Theoretische Physik, Universit\"at Giessen, Germany}

\author{L.~Alvarez-Ruso}
\affiliation{Departamento de F\'{\i}sica, Universidad de Murcia, Spain}

\date{December 9, 2008}

\begin{abstract}
  We investigate charged and neutral current neutrino induced incoherent pion production
  off nuclei at MiniBooNE and K2K energies within the GiBUU model. We assume impulse
  approximation and treat the nucleus as a local Fermi gas of nucleons bound in a
  mean-field potential. In-medium spectral functions are also taken into account. The
  outcome of the initial neutrino nucleon reaction undergoes complex hadronic final state
  interactions. We present results for neutral current $\pi^0$ and charged current $\pi^+$
  production and compare to MiniBooNE and K2K data.
\end{abstract}

\pacs{13.15.+g, 25.30.Pt}

\maketitle

\textbf{Introduction.}
A proper understanding of neutrino induced pion production is essential for the
interpretation of current neutrino oscillations experiments since it plays an important
role in neutrino flavor identification.  $\pi^0$ production events in neutral current (NC)
reactions are a source of background in $\nu_e$ appearance searches because they might be
misidentified as charge current (CC) $(\nu_e,e^-)$ interactions.  Similarly, CC induced
$\pi^{+,0}$ mesons are a background in $\nu_\mu$ disappearance experiments. As all of the
present oscillations experiments use nuclear targets, it is mandatory to consider final
state interactions (FSI), i.e., pion rescattering, with and without charge exchange, and
absorption in the nuclear medium. Next generation experiments such as T2K require the CC
$\pi^+$ (NC $\pi^0$) cross section to be known to 5\% (10\%) for the resulting error
on the oscillation parameters to be comparable to that from statistical uncertainties
\cite{AguilarArevalo:2006se}.

Neutrino induced pion production on nucleons up to energies of about 1.5 GeV is dominated
by the excitation and subsequent decay of the $\Delta (1232)$ resonance but, depending on
the channel, non-resonant pion production is not negligible; at higher energies, heavier
resonances become increasingly important~
\cite{Fogli:1979cz,Rein:1980wg,Sato:2003rq,Lalakulich:2006sw,Hernandez:2007qq,Hernandez:2007ej,Leitner:2008ue}.
For nucleons bound in a nucleus, the cross sections are modified due to Fermi motion,
Pauli blocking, mean-field potentials, and collisional broadening of the particles
\cite{Singh:1998ha,Leitner:2008ue}. Praet \refetal{Praet:2008yn} employ realistic
bound-state wave functions and find a good agreement to relativistic Fermi gas
calculations at neutrino energies around 1 GeV.

On nuclei, pions can be produced either coherently, leaving the nucleus intact, or
incoherently. While the former one has attracted considerable attention, the literature on
incoherent processes is limited. A full description of the pion production process
requires a realistic treatment of the FSI in the nucleus as outlined before.  The study of
semi-inclusive pion production presented in \refcite{Paschos:2000be} includes pion
absorption and charge exchange but does not properly take into account the important
features of $\pi N \Delta$ dynamics, leading to unrealistic pion spectra
\cite{Leitner:2007px,Leitner:2006sp}.  Extending the model of \refcite{Singh:1998ha},
Ahmad \refetal{Ahmad:2006cy} include, besides pion absorption, also elastic and charge
exchange rescattering using empirical vacuum $\pi N$ cross sections. FSI are neglected in
\refcite{Praet:2008yn} who calculate pion spectra in a plane wave impulse approximation.
 
Monte Carlo event generators, which are used in the simulation of the neutrino
experiments, are commonly based on the Rein-Sehgal model for pion production on the
nucleon \cite{Rein:1980wg} but they differ significantly in the treatment of nuclear
effects and FSI (cf.~\refcite{Andreopoulos:2007zz} and references therein). First
MiniBooNE and K2K results indicate a disagreement between the Monte Carlo predictions and
the actual measurements \cite{zellerneutrino2008Talk}.

In this brief report, we apply the Giessen Boltzmann-Uehling-Uhlenbeck (GiBUU) model for
neutrino induced reactions to NC and CC pion production on Carbon and Oxygen nuclei at K2K
and MiniBooNE energies. General results were presented in
\refscite{Leitner:2006ww,Leitner:2006sp}. The GiBUU model is based on well-founded
theoretical ingredients and has been tested in various and very different reactions, in
particular, against electron and photon scattering data
\cite{Krusche:2004uw,Leitner:2008ue}.

\textbf{GiBUU model.}  
In our model, neutrino nucleus scattering is treated as a two step
process. In the initial state step, the neutrinos interact with nucleons embedded in the
nuclear medium as explained in detail in \refcite{Leitner:2008ue}. In the final state
step, the outgoing particles of the initial reaction are propagated through the nucleus,
using a hadronic transport approach \cite{gibuu}. As we shall see, these FSI modify the
pion yields considerably.

We treat the nucleus as a local Fermi gas of nucleons bound in a mean-field potential and
obtain for the total pion production cross section on nuclei
\begin{multline}
  \dd \sigma (\nu_\ell A \to \ell' \pi X) = \\ \int \dd^3 r \int^{p_F(r)} \frac{\dd^3 p }{(2\pi)^3}  \frac{k \cdot p}{\kz \pz}  \dd \sigma_\subtot^{\text{med}} (\nu_\ell N \to \ell' X) M_\pi.
\end{multline}
$k$ is the four-vector of the the neutrino, $p$ the one of the bound nucleon, and
$p^{n,p}_F(r)=(3\pi^2\rho^{n,p}(r))^{1/3}$ denotes the local Fermi momentum depending on
the nuclear density. $M_\pi$ is the multiplicity of the final state which is determined by
the GiBUU transport simulation described below. $\dd \sigma_\subtot^{\text{med}}$ stands
for the total cross section on nucleons including nuclear medium corrections.

For neutrino beam energies ranging from 0.5 to 2 GeV, pions are produced both through
resonance and background contributions \cite{Leitner:2008ue}. Even initial quasielastic
scattering (QE) events contribute to the pion production cross section through secondary
collisions in the FSI. The pion production cross section is dominated by the excitation of
the $\Delta$ resonance \res{P}{33}{1232}.  Additionally, we include 12 $N^*$ and $\Delta$
resonances with invariant masses less than 2 GeV and also non-resonant pion production
(treated as background in our model) which is non-negligible in some cases.  The vector
parts of the resonance contributions are obtained from a recent MAID analysis of
electroproduction cross sections, the same holds for the background vector contributions.
The axial couplings for the resonances are obtained from PCAC (partial conservation of the
axial current) as described in \refcite{Leitner:2008ue}. We further use neutrino
nucleon data to fit the axial form factor of the $\Delta$ resonance as well as the non-vector
background contribution.  We refer the reader to \refcite{Leitner:2008ue} where our model
for lepton-nucleon scattering is described in detail.

The neutrino nucleon cross sections are modified in the nuclear medium. Bound nucleons are
treated within a local Thomas-Fermi approximation. They are exposed to a mean-field
potential depending on density and momentum. We account for this by evaluating the above
cross sections with full in-medium kinematics, i.e., hadronic tensor, flux and phase-space
factors are evaluated with in-medium four-vectors. We also take Pauli blocking into
account.

Once produced inside the nucleus, the particles propagate out to the detector. During
their propagation they undergo FSI which are simulated with the coupled-channel
semi-classical GiBUU transport model
\cite{gibuu}.\footnote{The numerical implementation of the GiBUU model is available for
  download from our website~\cite{gibuu}.}  Originally developed to
describe heavy-ion collisions, it has been extended to describe the interactions of pions,
photons, electrons, and neutrinos with
nuclei~\cite{Leitner:2008ue,Leitner:2006ww,Leitner:2006sp, Buss:2006vh}. 

In the following, we give a brief review on the basic ingredients of our model; for full
details we refer the reader to \refscite{gibuu} and references therein.  It is
based on the BUU equation which describes the space-time evolution of a many-particle
system in a mean-field potential. For particles of species $i$, it is given by
\begin{multline}
\left({\partial_t}+\myvec\nabla_p H\cdot\myvec\nabla_r  -\myvec\nabla_r H\cdot\myvec\nabla_p\right)f_i(\myvec{r},p, t) =  \\ I_{coll}[f_i,f_N,f_\pi,f_{\Delta},...],
\end{multline}
where the phase space density $f_i(\myvec{r},p,t)$ depends on time $t$, coordinates
$\myvec{r}$ and the four-momentum $p$.  $H$ is the relativistic Hamiltonian of a particle
of mass $M$ in a scalar potential $U$ given by $H=\sqrt{\left[ M +
    U(\myvec{r},\myvec{p})\right]^2 + \myvec{p}^{\,2} }$.
The collision term $I_{coll}$ accounts for changes (gain and loss) in the phase space
density due to elastic and inelastic collisions between the particles, and also to
particle decays into other hadrons. The BUU equations for all particle species are thus
coupled through the collision term and also through the potentials in $H$. A
coupled-channel treatment is required to take into account side-feeding into
different channels.  Baryon-meson two-body interactions (e.g., $\pi N \to \pi N$) are
dominated by resonance contributions and a small non-resonant background term;
baryon-baryon cross sections (e.g., $NN \to NN$, $R N \to N N$, $R N \to R' N$, $N N \to
\pi NN$) are either fitted to data or calculated e.g., in pion exchange models. The three-body
channels $\pi N N \to NN$ and $\Delta N N \to NNN$ are also included.

All particles (also resonances) are propagated in mean-field potentials according to their
BUU equation.  Those states acquire medium-modified spectral functions (nucleons and
resonances) and are propagated off-shell. 
The medium-modification of the spectral function is based both on collisional broadening and on the mean-field potentials. The 
collisional broadening is calculated using the low-density
approximation $\Gamma_\text{coll} = \sigma(E,\myvec{p},\myvec{p'}) v_\text{rel} \rho
(r)$, where $\sigma(E,\myvec{p},\myvec{p'})$ denotes the total cross section for the
scattering of the outgoing nucleon/resonance of energy $E$ and momentum $\myvec{p}$ with a
nucleon of momentum $\myvec{p'}$ in the vacuum with relative velocity $v_\text{rel}$. The
collisional broadening is obtained in a consistent way from the GiBUU cross sections, and
depends on the particle kinematics as well as on the nuclear density. In
\refcite{Buss:2007ar,Leitner:2008ue}, we have shown that both the momentum dependence of
the mean-field potential and the collisional broadening are necessary to obtain good
agreement with inclusive electron scattering experimental data.  We ensure that after leaving the
nucleus, vacuum spectral functions are recovered. Finally, the pion multiplicity $M_\pi$
is determined by counting all asymptotic pions in each kinematical bin.

Summarizing, FSI lead to absorption, charge exchange and redistribution of energy and
momentum, as well as to the production of new particles. Their impact on neutrino induced
pion production is dramatic \cite{Leitner:2006ww,Leitner:2006sp} and, therefore, a
qualitatively and quantitatively correct treatment is of great importance.

\begin{figure}
  \centering
  \includegraphics[scale=\plotscale]{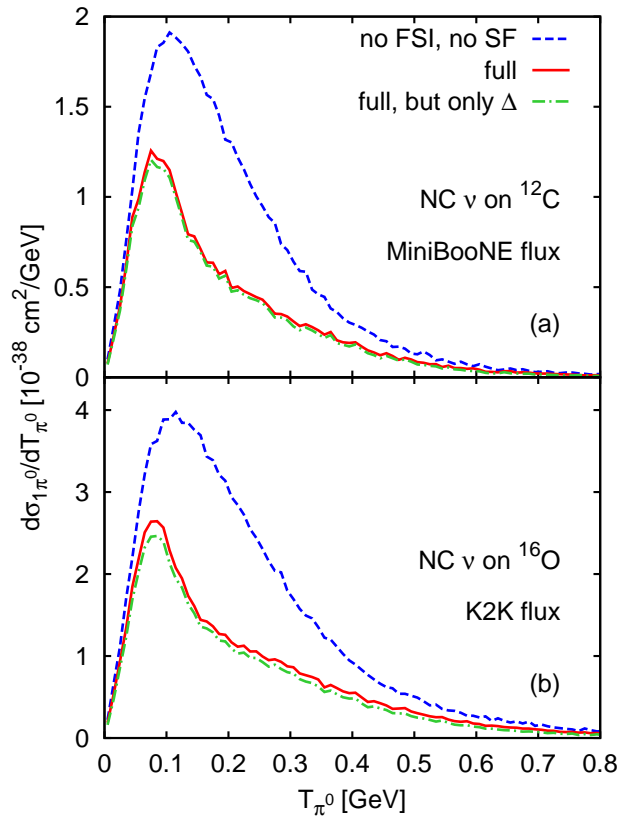}
  \vspace{-1ex}
  \caption{(Color online) (a) NC induced single-$\pi^0$ production on $^{12}$C as a function of
    the pion kinetic energy averaged over the MiniBooNE flux. (b) same on
    $^{16}$O averaged over the K2K flux. The dashed line shows our calculation without FSI
    or spectral functions, both included in the full calculation denoted by
    the solid line. The dash-dotted line indicates the contribution from the $\Delta$
    resonance to the full calculation.}
  \label{fig:MiniBooNE_K2K_NC}
\end{figure}

\textbf{Results.}  In \reffig{fig:MiniBooNE_K2K_NC} (a), we show our results for
NC single-$\pi^0$ production on $^{12}$C as a function of the pion kinetic energy. We have
averaged over the MiniBooNE energy flux which peaks at about 0.7 GeV neutrino energy
\cite{AguilarArevalo:2008yp}.  As can be deduced from simple isospin arguments, in NC
reactions the total pion yield is dominated by $\pi^0$ production, while $\pi^+$ dominate
in CC processes.  Comparing the dashed with the solid line (results without FSI and
spectral function vs.~full calculation), one finds a considerably change of the spectra.
The shape is caused by the energy dependence of the pion absorption and rescattering cross
sections. Pions are mainly absorbed via the $\Delta$ resonance, i.e, through $\pi N \to
\Delta$ followed by $\Delta N \to NN$. This explains the reduction in the region around
$T_\pi=0.1-0.3\GeV$.  Pion elastic scattering $\pi N \to \pi N$ reshuffles the pions to
lower momenta and leads also to charge exchange scattering into the charged pion channels.
The vast majority of the pions come from initial $\Delta$ excitation (dash-dotted line).
Their production in the rescattering of nucleons is not
significant here  but becomes more important at higher energies and for heavier
nuclei \cite{Leitner:2006sp}.

The MiniBooNE experiment has recently measured NC single-$\pi^0$ momentum spectra
\cite{AguilarArevalo:2008xs}, however, their data are available only as count rates.
Notice that the data include a small contribution from coherent pion production, i.e., $\nu A
\to \nu \pi^0 A$, which cannot be described by our transport model. A direct
and meaningful comparison to the MiniBooNE measurement will be possible when acceptance 
corrected cross sections are provided.

\reffig{fig:MiniBooNE_K2K_NC} (b) shows the results for NC single-$\pi^0$
production off $^{16}$O averaged over the K2K energy flux which peaks at about 1.2 GeV
neutrino energy \cite{Nakayama:2004dp}. Compared to (a), the spectrum is
broader and extends to larger $T_\pi$ due to the higher neutrino energy. The reduction in
the region around $T_\pi=0.1-0.3\GeV$ is mainly caused by the pion absorption via the
$\Delta$ resonance (compare dashed and solid lines). Again, pion production through
initial QE scattering is not sizable.

NC single-$\pi^0$ spectra have been measured by the K2K collaboration
\cite{Nakayama:2004dp}. Based on their Monte Carlo generator, the data are not only
corrected for efficiency, but some background has already been subtracted, i.e., the data
include a model dependence. Since K2K, as well as MiniBooNE, has not yet provided cross
sections but only count rates, we cannot yet compare to these measurements directly.
 
\begin{figure}
  \centering
  \includegraphics[scale=\plotscale]{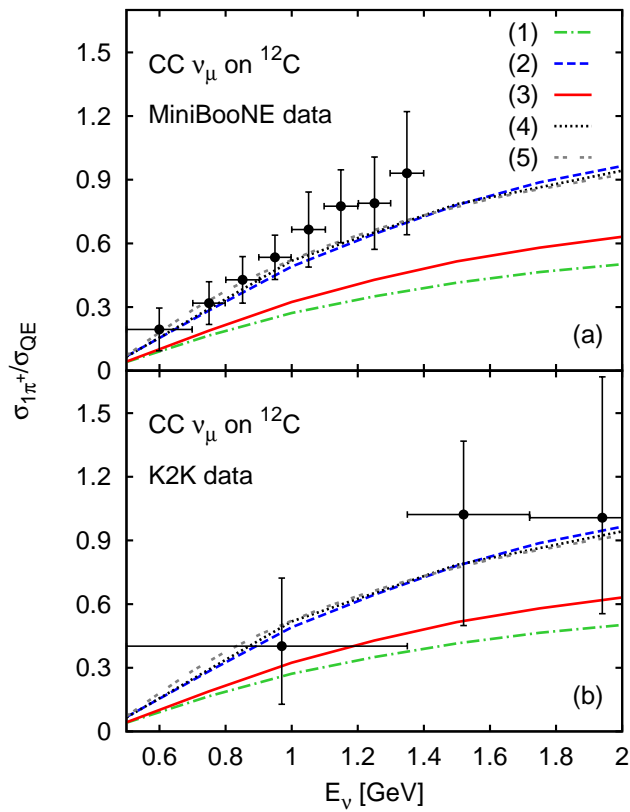} 
  \vspace{-1ex}
  \caption{(Color online) Single-$\pi^+$/QE cross section ratio for CC interactions on
    $^{12}$C vs.~neutrino energy. Our results are the same in both panels. The meaning of
    the five different curves is described in the text. (a) shows preliminary data from
    MiniBooNE \cite{Wascko:2006tx}; (b) recent K2K data \cite{Rodriguez:2008ea}. The
    curves (2), (4), and (5) lie nearly on top of each other.}
  \label{fig:MB_K2K_CC}
\end{figure}

In \reffig{fig:MB_K2K_CC}, we give our results for the single-$\pi^+$/QE ratio for CC
interactions on $^{12}$C. We present different scenarios:
\begin{enumerate}
\setlength{\itemsep}{-1ex}
\item $\sigma_{1\pi^+} / \sigma_{0\pi^+}$ after FSI: This denotes our full calculation
  after FSI for the single-$\pi^+$ cross section divided by the ``CCQE-like'' cross
  section, i.e., the cross section after FSI without any pion leaving the nucleus. This
  definition of ``CCQE-like'' as all the events where no pion is detected, is used,
  e.g., by MiniBooNE.
\item $\sigma_{1\pi^+} / \sigma_{0\pi^+\,1 p}$ after FSI: similar to (1), but we require
  in addition a single-$p$ in the final state for the ``CCQE-like'' cross section. Here,
  ``CCQE-like'' denotes all events where a single proton track is visible and, at the same
  time, no pions are detected as applied, e.g., by K2K.
\item $\sigma_{1\pi^+}$ after FSI$/ \sigma_{\text{QE}}$: Here, the full single-$\pi^+$
  cross section after FSI is divided by the total QE cross section.
\item $\sigma_{1\pi^+}$ before FSI$/ \sigma_{\text{QE}}$: Same as (3), but before FSI;
  nuclear corrections like mean fields and Fermi motion are included.
\item $\sigma_{1\pi^+} / \sigma_{\text{QE}}$ in the vacuum: For comparison, we show
  the vacuum cross section on an isoscalar target.
\end{enumerate}
In (3)-(5) $\sigma_{\text{QE}}$ defines the situation in which the primary
neutrino-nucleon interaction is purely quasielastic ($\nu_\mu n \rightarrow \mu^- p$) regardless
of the fate of the struck nucleon.

We want to emphasize two issues: First, nuclear corrections cancel out in the
ratio, as long as FSI are not considered ((4) vs.~(5)). In general, the complexity of FSI
prevent such cancellations as one can infer from (1) and (3). Only on very specific
occasions, FSI effects may cancel: (2) with the particular definition of the ``CCQE-like''
cross section ($=\sigma_{0\pi^+\,1 p}$) lies nearly on top of (4) and (5).

We further compare to preliminary MiniBooNE data \cite{Wascko:2006tx}
(\reffig{fig:MB_K2K_CC} (a)) and to K2K data \cite{Rodriguez:2008ea}
(\reffig{fig:MB_K2K_CC} (b)). These data are corrected for FSI using specific Monte Carlo
generators, i.e., they give the cross sections for bound nucleons ``before FSI''. As this
procedure introduces a model dependence in the data, a fully consistent comparison is not
possible. Furthermore, also the reconstruction of the neutrino energy out of the observed
muon and the hadrons is model dependent. Ignoring the model dependencies, our calculation
denoted by (4) should be the one to compare with. In the MiniBooNE case, the agreement is
perfect for energies up to 1 GeV, and still within their error bars above 1 GeV (cf.~\reffig{fig:MB_K2K_CC} (a)). We also reach a good agreement with the K2K data
(cf.~\reffig{fig:MB_K2K_CC} (b)). The slight underestimate of the
pion/quasielastic ratio at higher energies could be due to either an underestimate of the
pion production cross section, an overestimate of the QE cross section, or to
uncertainties in the ``data'' extraction which involves a model dependence (see above).

To summarize, in this brief report, we have presented results for NC and CC single-$\pi$
production for MiniBooNE and K2K energies. In particular, we have investigated the effect
of final state interactions. Wherever possible, we have compared our
calculation to recent data and we have found good agreement. However, all these ``data''
were readjusted using specific Monte Carlo event generator with specific assumptions
on the initial neutrino nucleon cross section and the nuclear model. Model independent
acceptance corrected data are required to perform meaningful comparisons with theoretical
calculations.

\begin{acknowledgments}
  We thank O. Lalakulich for fruitful discussions. This work has been supported by the
  Deutsche Forschungsgemeinschaft. LAR acknowledges financial support from the Seneca
  Foundation.
\end{acknowledgments}

\vspace{-4ex}


\begin{thebibliography}{10}

\bibitem{AguilarArevalo:2006se}
A.~A. Aguilar-Arevalo {\em et~al.},
\newblock arXiv:hep-ex/0601022.

\bibitem{Fogli:1979cz}
G.~L. Fogli and G.~Nardulli,
\newblock Nucl. Phys. B {\bf 160}, 116 (1979).

\bibitem{Leitner:2008ue}
T.~Leitner, O.~Buss, L.~Alvarez-Ruso and U.~Mosel,
\newblock Phys. Rev. C {\bf 79}, 034601 (2009).

\bibitem{Rein:1980wg}
D.~Rein and L.~M. Sehgal,
\newblock Ann. Phys. {\bf 133}, 79 (1981).

\bibitem{Sato:2003rq}
T.~Sato, D.~Uno and T.~S.~H. Lee,
\newblock Phys. Rev. C {\bf 67}, 065201 (2003).

\bibitem{Hernandez:2007qq}
E.~Hernandez, J.~Nieves and M.~Valverde,
\newblock Phys. Rev. D {\bf 76}, 033005 (2007).

\bibitem{Hernandez:2007ej}
E.~Hernandez, J.~Nieves, S.~K. Singh, M.~Valverde and M.~J. Vicente~Vacas,
\newblock Phys. Rev. D {\bf 77}, 053009 (2008).

\bibitem{Lalakulich:2006sw}
O.~Lalakulich, E.~A. Paschos and G.~Piranishvili,
\newblock Phys. Rev. D {\bf 74}, 014009 (2006).

\bibitem{Singh:1998ha}
S.~K. Singh, M.~J. Vicente-Vacas and E.~Oset,
\newblock Phys. Lett. B {\bf 416}, 23 (1998).

\bibitem{Praet:2008yn}
C.~Praet, O.~Lalakulich, N.~Jachowicz and J.~Ryckebusch,
\newblock arXiv:0804.2750.

\bibitem{Paschos:2000be}
E.~A. Paschos, L.~Pasquali and J.-Y. Yu,
\newblock Nucl. Phys. B {\bf 588}, 263 (2000).

\bibitem{Leitner:2007px}
T.~Leitner, O.~Buss, U.~Mosel and L.~Alvarez-Ruso,
\newblock AIP Conf. Proc. {\bf 967}, 192 (2007).

\bibitem{Leitner:2006sp}
T.~Leitner, L.~Alvarez-Ruso and U.~Mosel,
\newblock Phys. Rev. C {\bf 74}, 065502 (2006).

\bibitem{Ahmad:2006cy}
S.~Ahmad, M.~Sajjad~Athar and S.~K. Singh,
\newblock Phys. Rev. D {\bf 74}, 073008 (2006).

\bibitem{Andreopoulos:2007zz}
C.~Andreopoulos,
\newblock AIP Conf. Proc. {\bf 967}, 31 (2007).

\bibitem{zellerneutrino2008Talk} 
G.~Zeller, 
\newblock talk given at Neutrino 08, May 25-31,
2008, Christchurch, New Zealand.

\bibitem{Leitner:2006ww}
T.~Leitner, L.~Alvarez-Ruso and U.~Mosel,
\newblock Phys. Rev. C {\bf 73}, 065502 (2006).

\bibitem{Krusche:2004uw}
B.~Krusche {\em et~al.},
\newblock Eur. Phys. J. {\bf A22}, 277 (2004).

\bibitem{gibuu}
Gi{BUU},
\newblock \url{http://gibuu.physik.uni-giessen.de/GiBUU}.

\bibitem{Buss:2006vh}
O.~Buss, L.~Alvarez-Ruso, P.~Muehlich and U.~Mosel,
\newblock Eur. Phys. J. {\bf A29 (2)}, 189 (2006).

\bibitem{Buss:2007ar}
O.~Buss, T.~Leitner, U.~Mosel and L.~Alvarez-Ruso,
\newblock Phys. Rev. C {\bf 76}, 035502 (2007).

\bibitem{AguilarArevalo:2008yp}
A.~A. Aguilar-Arevalo {\em et~al.},
\newblock arXiv:0806.1449.

\bibitem{AguilarArevalo:2008xs}
A.~A. Aguilar-Arevalo {\em et~al.},
\newblock Phys. Lett. B {\bf 664}, 41 (2008).

\bibitem{Nakayama:2004dp}
S.~Nakayama {\em et~al.},
\newblock Phys. Lett. B {\bf 619}, 255 (2005).

\bibitem{Wascko:2006tx}
M.~O. Wascko,
\newblock Nucl. Phys. Proc. Suppl. {\bf 159}, 50 (2006).

\bibitem{Rodriguez:2008ea}
A.~Rodriguez {\em et~al.},
\newblock Phys. Rev. D {\bf 78}, 032003 (2008).

\end{thebibliography}

\end{document}